\documentclass[journal,twoside]{IEEEtran}
\usepackage{cite}
\usepackage{comment}
\usepackage{epsfig}
\usepackage{graphicx}
\usepackage{subfigure}
\usepackage{amsmath,amssymb}
\usepackage{color}

\newcommand{\beq}{\begin{equation}}
\newcommand{\eeq}{\end{equation}}
\newcommand{\beqr}{\begin{eqnarray}}
\newcommand{\eeqr}{\end{eqnarray}}

\newcommand{\E}{\mathbb{E}}
\newcommand{\ns}{\widetilde{Y}_{k,j}^{(m)}}
\newcommand{\F}{F}
\newcommand{\W}{W}
\newcommand{\dz}{Z}

\begin{document}

\title{Adaptive State-Space Multitaper Spectral Estimation}

\author{Andrew H. Song$^{\ast}$,~\IEEEmembership{Student Member,~IEEE,} Seong-Eun~Kim$^{\ast}$,~\IEEEmembership{Member,~IEEE,} and Emery N. Brown,~\IEEEmembership{Fellow,~IEEE} %
\thanks{$^\ast$ These authors contributed equally to this work.}
\thanks{This work was partially supported by NRF grants (2020R1C1C1011857, 2020M3C1B8081320, 2019S1A5A2A03053308), and by NIH
grant P01-GM118629 and the JPB Foundation \textit{(Corresponding author: Seong-Eun Kim)}}
\thanks{A.~H.~Song is with Electrical Engineering and Computer Science, Massachusetts Institute of Technology, Cambridge, MA 02139, USA (e-mail:andrew90@mit.edu).}
\thanks{S.-E.~Kim is with the Department of Applied Artificial Intelligence, Seoul National University of Science and Technology, Nowon-gu, Seoul 01811, South Korea (e-mail:sekim@seoultech.ac.kr).}
\thanks{E.~N.~Brown is with the Picower Institute for Learning and Memory, Massachusetts Institute of Technology, Cambridge, MA 02139 and the Department of Anesthesia, Critical Care, and Pain Medicine, Massachusetts General Hospital, Harvard Medical School, Boston, MA 02114, USA (e-mail:enb@neurostat.mit.edu).}}



\maketitle

\begin{abstract}
Short-time Fourier transform (STFT) is the most common window-based approach for analyzing the spectrotemporal dynamics of time series. To mitigate the effects of high variance on the spectral estimates due to finite-length, independent STFT windows, state-space multitaper (SSMT) method used a state-space framework to introduce dependency among the spectral estimates. However, the assumed time-invariance of the state-space parameters makes the spectral dynamics difficult to capture when the time series is highly nonstationary. We propose an adaptive SSMT (ASSMT) method as a time-varying extension of SSMT. ASSMT tracks highly nonstationary dynamics by adaptively updating the state parameters and Kalman gains using a heuristic, computationally efficient exponential smoothing technique. In analyses of simulated data and real human electroencephalogram (EEG) recordings, ASSMT showed improved denoising and smoothing properties relative to standard multitaper and SSMT approaches.

\end{abstract}

\textit{This manuscript is an extended version of the IEEE Signal
Processing Letters paper (doi:10.1109/LSP.2022.3142670), with the supplementary material as the appendix.}\\


\IEEEpeerreviewmaketitle

\section{Introduction}
\IEEEPARstart{N}{onstationary} time series with time-varying probability structures are ubiquitous. Some examples include speech and image recordings \cite{speech,image}, oceanographic and seismic signals \cite{ocean}, neural spike trains \cite{spike}, and electroencephalogram (EEG) \cite{EEG}. We are interested in analyzing the nonstationary data through the lens of the time-varying spectral dynamics, which yields valuable information on the underlying system. The traditional approach has been to segment the data into \textit{independent} overlapping or non-overlapping windows, assuming local stationarity~\cite{Dahlhaus97} within each window, and to apply Fourier or wavelet transform~\cite{wavelet, Oppenheim}.

Despite the popularity, the windowing approach suffers from spectral estimates of high variance within each window, due to finite window-length~\cite{Fourier} and the restrictive independence assumption for different windows. The state-space multitaper (SSMT) framework~\cite{SSMT} and its extensions \cite{Ba, Das, Song18, Song21} have been proposed as the solutions, by positing a \textit{time-invariant} latent state-space model in the time-frequency domain, with each state representing the Fourier coefficients in each window. Since the states are linked by stochastic continuity constraint across windows, the spectral estimates are not independent. 

However, the use of time-invariant parameters limits the capacity of SSMT to track strong nonstationarity that commonly occurs in systems neuroscience and manifests as strong spectral fluctuations due to fluctuations in system properties. Although time-varying state-space model offers an obvious solution, traditional parameter estimation approaches based on expectation-maximization (EM) algorithm~\cite{Dempster77} makes it less practical for real-time applications - The parameters must be re-estimated with every incoming batch of data, and thus incurs high computational costs.



We propose a time-varying extension of SSMT with an efficient and adaptive parameter estimation scheme,
termed an adaptive SSMT (ASSMT) method. Based on a novel nonstationarity metric derived from spectral fluctuation, ASSMT adaptively adjusts the state parameters and consequently Kalman gain. This allows ASSMT to switch between removing background noise and tracking spectral fluctuation in a data-driven manner. The estimation for time-varying parameters requires only single pass through the data, making it well suited for real-time applications. The paper is organized as follows.
In Section II, we review the SSMT framework. In Section III, we formulate the ASSMT framework. In Section IV and V, the results and conclusion are presented.

\section{Review of State-Space Multitaper Method}
We first review the SSMT algorithm in \cite{SSMT}. Consider a nonstationary time series $y_t$ sampled at frequency $f_s$ as
\begin{align} y_t=x_t+\varepsilon_t,\quad t=1,\ldots, T \end{align}
where $x_t$ is a locally stationary latent Gaussian process~\cite{Adak98} with the measurement noise $\varepsilon_t\sim\mathcal{N}(0,\sigma_{\varepsilon}^2)$. Leveraging the local stationary property, we divide these signals into $K$ nonoverlapping stationary intervals of $J$ samples, such that $T=KJ$. The segmented vectors for interval $k$ are denoted as $Y_k, X_k, \varepsilon_k\in\mathbb{R}^J$, with the $j$th element as $Y_{k,j}=y_{J(k-1)+j}$ for $k=1,\ldots,K$ and $j=1,\ldots,J$. 

To perform time-frequency analysis, we introduce the latent $\dz_k=(\dz_{k,1},\ldots,\dz_{k,J})\in\mathbb{C}^J$, where $\dz_{k,j}$ is a complex Gaussian variable with the magnitude corresponding to the power at the normalized frequency $\omega_j=2\pi(j-1)/J$ and interval $k$~\cite{Brillinger}. To model the evolution of spectra across the windows, we assume that $\{\dz_k\}_{k=1}^K$ follow a random walk prior
\begin{align} \dz_k = \dz_{k-1}+v_k,\end{align}
where the state noise $v_k$ is a complex Gaussian process with a diagonal covariance $I(\sigma^2_{v,j})$, i.e., $v_k\sim\mathcal{CN}(0, I(\sigma^2_{v,j}))$. The prior encodes two important properties of $\{\dz_k\}_{k=1}^K$. First, the state variance $\sigma_{v,j}^2$ controls the smoothness of the process, with a large value indicative of non-smooth or fluctuating process. Second, $\dz_{k,j}$ and $\dz_{k,j'}$ for $j\neq j'$ are independent \textit{a priori}. 

We can link the time-frequency process $\{\dz_k\}_{k=1}^K$ with time-domain observation $\{Y_k\}_{k=1}^K$, 
using the Fourier transform matrix $\F\in\mathbb{C}^{J\times J}$ with $F_{j,l}=J^{-1/2}\exp(-i2\pi(l-1)j/J)$ and the inverse Fourier matrix $\W\in\mathbb{C}^{J\times J}$, such that $\F\W=I$,
\begin{align} 
&Y_k=X_k+\varepsilon_k=\W\dz_k+\varepsilon_k\\
&\Rightarrow \F Y_k=\F\W\dz_k+\F\varepsilon_k=\dz_k+\varepsilon_k^F,\nonumber
\end{align}
where $\varepsilon_k^F\sim\mathcal{CN}(0,I(\sigma_{\varepsilon}^2))$, since $FI(\sigma_{\varepsilon}^2)W=I(\sigma_{\varepsilon}^2)$.

Along with the state-space model, we incorporate the data tapers to further reduce the variance of the spectral estimates. Specifically, we use $M$ Slepian tapers, leading to the multitaper (MT) method that optimally balances the bias-variance trade-off via bandwidth adjustment \cite{Thomson82, MT}. This essentially produces $M$ independent set of state-space models,
\begin{align}
    Y^{(m),F}_k &= \dz^{(m)}_k+\varepsilon^{(m),F}_k \label{eq:obs_model} \\
    \dz_k^{(m)} &= \dz_{k-1}^{(m)}+v^{(m)}_k \label{eq:ss_model},
\end{align}
where $v_k^{(m)}\sim\mathcal{CN}(0, I(\sigma^{2,(m)}_{v,j}))$, $\varepsilon_k^{(m),F}\sim\mathcal{CN}(0,I(\sigma_{\varepsilon}^{2,(m)}))$, $Y_k^{(m)}$ denotes the $m^{\text{th}}$ Slepian taper applied to $Y_k$, $Y_k^{(m),F}$ denotes Fourier transform of $Y_k^{(m)}$, i.e., $Y_k^{(m),F}=\F Y_k^{(m)}$, and $\dz^{(m)}_k$ represents the $m$th spectral eigen-coefficient of $\dz_k$. The variants of SSMT modify Eqs.~(\ref{eq:obs_model}) or (\ref{eq:ss_model}) \cite{Ba, Das, Song18, Song21}.

We now focus on $\omega_j$ for simplicity. Based on Eqs.~(\ref{eq:obs_model}) and (\ref{eq:ss_model}), we derive a Kalman filter algorithm for the estimation of $\dz^{(m)}_{k,j}$ using Kalman gain $C^{(m)}_{k,j}$
\begin{align}
    \dz^{(m)}_{k|k,j}&=(1-C^{(m)}_{k,j})\dz^{(m)}_{k-1|k-1,j}+C^{(m)}_{k,j}Y^{(m),F}_{k,j}\label{eq:filter} \\
    \sigma^{2,(m)}_{k|k,j}&=(1-C^{(m)}_{k,j})(\sigma^{2,(m)}_{k-1|k-1,j}+\sigma^{2,(m)}_{v,j}),
\end{align}
where the notation $k|s$ denotes the estimate on interval $k$ given the data observed up to interval $s$ and $C^{(m)}_{k,j}$ is given as
\begin{align}
    C^{(m)}_{k,j}=\frac{\sigma^{2,(m)}_{k-1|k-1,j}+\sigma^{2,(m)}_{v,j}}{\sigma_\varepsilon^{2,(m)}+\sigma^{2,(m)}_{k-1|k-1,j}+\sigma^{2,(m)}_{v,j}}.\label{eq:kg}
\end{align}
The spectrogram estimate at frequency $\omega_j$ on interval $k$ is
\begin{align} 
f^{\text{SSMT}}_{k}(\omega_j)=M^{-1}\sum^{M}_{m=1}|\dz^{(m)}_{k|k,j}|^2. 
\end{align}
The parameters $\{\sigma_{v,j}^{2,(m)}\}_{j,m=1}^{J,M}$ and $\{\sigma_{\varepsilon}^{2,(m)}\}_{m=1}^M$ are estimated with EM algorithm~\cite{Dempster77}. SSMT tracks nonstationarity in the data with the \textit{time-invariant} parameters $\sigma^{2,(m)}_{v,j}$ and $\sigma_{\varepsilon}^{2,(m)}$, since $\dz^{(m)}_{k|k,j}\neq \dz^{(m)}_{k-1|k-1,j}$ and thus $f^{\text{SSMT}}_{k}(\omega_j)\neq f^{\text{SSMT}}_{k-1}(\omega_j)$.

\section{Adaptive SSMT}
\begin{figure}[!ht]
\centering
\includegraphics[width=\linewidth]{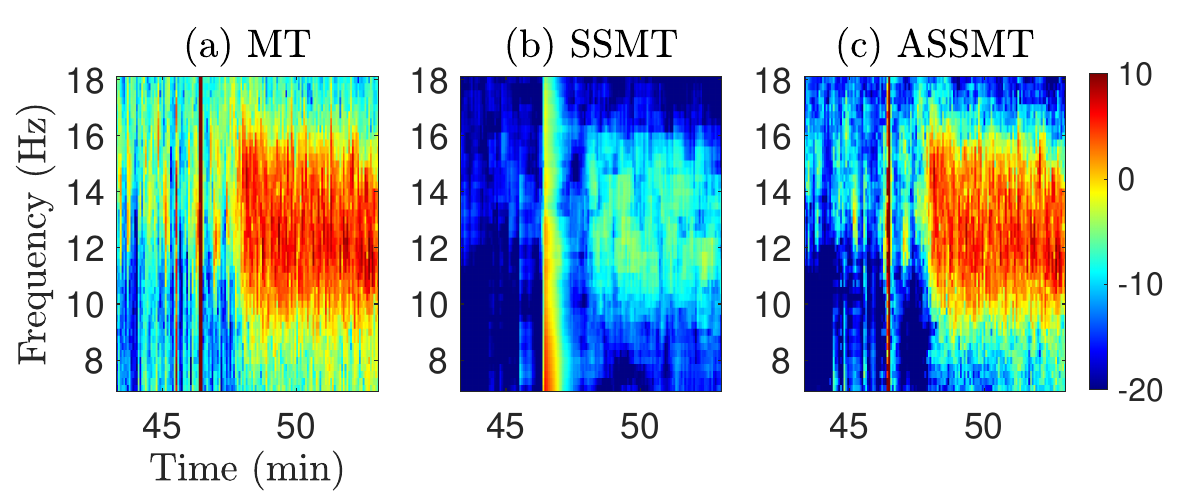}
  \caption{A spectrogram snippet estimated with (a) MT (b) SSMT (c) ASSMT.}
  \label{fig:comparison}
\end{figure}

Although SSMT can model slowly time-varying spectral dynamics, it is restrictive for highly fluctuating spectral dynamics. Such strong nonstationarity (or fluctuation) is common in EEG with \textit{external intervention} to the brain or with the change of the brain state during anesthesia/sleep~\cite{Purdon}. Fig.~\ref{fig:comparison} shows a snippet of human anesthesia EEG in which SSMT (Fig.~\ref{fig:comparison}(b)) cannot track the apparent dynamics shown by the MT approach (Fig.~\ref{fig:comparison}(a)). However, the proposed ASSMT (Fig.~\ref{fig:comparison}(c)) is able to capture the spectral dynamics and remove non-relevant background activity.

The failure of SSMT is due to the time-invariant parameters, as the Kalman gain quickly converges to a steady-state value $C_{\infty,j}^{(m)}$~\cite{shumway}. Denoting the observation prediction error as $\Delta Y^{(m)}_{k,j}= Y^{(m),F}_{k,j} - Z_{k-1|k-1,j}^{(m),F}$ and the latent estimate change as $\Delta Z_{k,j}^{(m)}=\dz^{(m)}_{k|k,j}-\dz^{(m)}_{k-1|k-1,j}$, we can express Eq.~(\ref{eq:filter}) as $\Delta Z_{k,j}^{(m)}=C^{(m)}_{\infty,j}\Delta Y_{k,j}^{(m)}$. If $C_{\infty,j}^{(m)}$ is low (SSMT for Fig.~\ref{fig:comparison}(b)), $\dz^{(m)}_{k|k,j}$ fails to reflect fluctuation in $Y_{k,j}^{(m),F}$. Consequently, $f_k^{\text{SSMT}}$ cannot reliably track the spectral dynamics.

To resolve this issue, adaptive SSMT (ASSMT) posits a time-varying state-space model, to allow the adaptive change of $\sigma_{v,j}^{2,(m)}$. Specifically, we replace $v_k^{(m)}\sim\mathcal{CN}(0, I(\sigma^{2,(m)}_{v,j}))$ in Eq.~(\ref{eq:ss_model}) with $v^{(m)}_{k,j}\sim\mathcal{CN}(0, I(\sigma^{2,(m)}_{v,k,j}))$, to indicate the state variance's dependence on time. This prevents Kalman gain from converging to a steady-state value, allowing the algorithm to produce more flexible spectrogram estimates. $\sigma_{\varepsilon}^{2,(m)}$ remains constant across windows because we assume stationary background noise.

With the modified generative model, we now address \textit{when} and \textit{how} ASSMT tracks varying degrees of nonstationarity. We first quantify the notion of nonstationarity, and then propose an adaptive parameter estimation approach.

\subsection{Measure of nonstationarity}
We define $\ns=\E|Y_{k,j}^{(m),F}-Y_{k-1,j}^{(m),F}|^2$ as the measure of nonstationarity, which is the expected observation difference and a function of window, frequency, and taper. Intuitively, we use high spectral fluctuation as a proxy for high nonstationarity, that is, when $\ns$ exceeds a frequency-dependent threshold $\beta_j$.

To estimate $\ns$, we use \textit{exponential moving average} (EMA), often used as a simple, yet effective approach to estimate the expectation in the filtering literature~\cite{adaptive_lms}
\begin{align}\label{eq:ema}
\ns =  (1-\alpha)\widetilde{Y}_{k-1,j}^{(m)} + \alpha |Y_{k,j}^{(m),F}-Y_{k-1,j}^{(m),F}|^2,
\end{align}
where $0 \leq \alpha \leq 1$ is a smoothing factor chosen \textit{a priori}. This approach allows us to use $\ns$ as a heuristic indicator of strong nonstationarity, decoupled from the generative model and hence the estimation procedure of $\dz^{(m)}_{k,j}$. The choice of $\alpha$ reflects the belief on the impulsiveness of nonstationarity and the volatile nature of the state transitions. With large $\alpha$, $\ns$ is sensitive to instantaneous fluctuation.

\subsection{Estimation of parameters \& adaptive thresholding}

We now examine how to use $\ns$ to set $\beta_j$ and subsequently estimate the parameters. Using Eq.~(\ref{eq:obs_model}), we have
\begin{align}\label{eq:diff_1}
&\ns=\E|Y_{k,j}^{(m),F}-Y_{k-1,j}^{(m),F}|^2\\
&\quad = \E|(\dz^{(m)}_{k,j}-\dz^{(m)}_{k-1,j})+ (\varepsilon^{(m),F}_{k}-\varepsilon^{(m),F}_{k-1})|^2\nonumber\\
&\quad=\E|\dz^{(m)}_{k,j}-\dz^{(m)}_{k-1,j}|^2+ \E|\varepsilon^{(m),F}_{k}-\varepsilon^{(m),F}_{k-1}|^2,\nonumber
\end{align}
where we use the uncorrelatedness of the two differences. Next, we use the fact that 1) $\E|\dz^{(m)}_{k,j}-\dz^{(m)}_{k-1,j}|^2=\sigma^{2,(m)}_{v,k,j}$ and 2) $\varepsilon^{(m),F}_{k}$ and $\varepsilon^{(m),F}_{k-1}$ are independent with variance $\sigma^{2,(m)}_\varepsilon$, hence $\E|\varepsilon^{(m),F}_{k}-\varepsilon^{(m),F}_{k-1}|^2=2\sigma_{\varepsilon}^{2,(m)}$, which leads to
\begin{align}\label{eq:sigma_relationship}
\E|Y_{k,j}^{(m),F}-Y_{k-1,j}^{(m),F}|^2= \sigma^{2,(m)}_{v,k,j} + 2\sigma^{2,(m)}_{\varepsilon}.
\end{align}
This establishes the connection between the nonstationarity metric, $\ns$, and the two sources of variance in our model.

With Eq.~(\ref{eq:ema}) and Eq.~(\ref{eq:sigma_relationship}), we can estimate $\sigma^{2,(m)}_{v,k,j}$. For $\sigma_{\varepsilon}^{2,(m)}$, we use $\widehat{\sigma}^{2,(m)}_{\varepsilon}$ estimated from SSMT. This leads to
\begin{align}\label{eq:ASS_MT_sv_original}
\widehat{\sigma}^{2,(m)}_{v,k,j} =  \ns - 2\widehat\sigma^{2,(m)}_{\varepsilon}.
\end{align}
We further lower bound $\widehat{\sigma}^{2,(m)}_{v,k,j}$ for two reasons. First, we impose that the time-varying SNR, $\gamma_{k,j}=\sigma^{2,(m)}_{v,k,j}/\sigma_{\varepsilon}^{2,(m)}$, is greater than the baseline SNR of the system, i.e., $\gamma_{k,j}\geq\gamma_{k,j}^{\text{baseline}}$. Second, we require nonnegative $\widehat{\sigma}^{2,(m)}_{v,k,j}$. Since SSMT estimates the baseline properties of the data, we set $\gamma_{k,j}^{\text{baseline}}=\widehat{\sigma}^{2,(m),\text{SSMT}}_{v,j}/\widehat{\sigma}_{\varepsilon}^{2,(m)}$. This yields the estimation procedure
\begin{align}\label{eq:ASSMT_sv}
    \widehat{\sigma}^{2,(m)}_{v,k,j} =  \max(\,\ns - 2\widehat\sigma^{2,(m)}_{\varepsilon},\, \widehat{\sigma}_{v,j}^{2,(m),\text{SSMT}}\,).
\end{align}

This obviates the need for EM beyond the initial phase for estimating $\widehat{\sigma}_{\varepsilon}^{2,(m)}$ and $\widehat{\sigma}_{v,j}^{2,(m),\text{SSMT}}$. Although EM can estimate the time-varying parameters, it requires multiple forward/backward passes through the entire data, which is computationally expensive. In addition, with every new observation, $\widehat{\sigma}^{2,(m)}_{v,k,j}$ in the earlier windows need to be re-estimated. 

\subsection{Estimation of nonstationary spectra}\label{section:kg}
We use Kalman filter with $\widehat{\sigma}^{2,(m)}_{v,k,j}$ to estimate the spectrogram $f^{\text{ASSMT}}_{k}(\omega_j)$. ASSMT thus operates with two different modes, depending on $\beta_j=2\widehat\sigma^{2,(m)}_{\varepsilon}+\widehat{\sigma}_{v,j}^{2,(m),\text{SSMT}}$. 
If $\ns \geq\beta_j$, ASSMT uses larger state variance to track high nonstationarity. Given $\sigma^{2,(m)}_{k-1|k-1,j}$ and $\widehat{\sigma}_{\varepsilon}^{2,(m)}$ in Eq.~(\ref{eq:kg}), we observe that the \textit{increase} in $\widehat{\sigma}^{2,(m)}_{v, k, j}$ leads to the \textit{increase} in $C_{k,j}^{(m)}$. This agrees with our intuition, since we want the Kalman gain to \textit{increase} such that $\Delta Z_{k,j}^{(m)}$ explains a greater portion of $\Delta Y^{(m)}_{k,j}$. For $\ns <\beta_j$, ASSMT simply uses the baseline $\widehat{\sigma}_{v,j}^{2,(m),\text{SSMT}}$ and thus operates with a fixed Kalman gain. This explains how ASSMT with adaptive state variance is able to track strong nonstationarity.

\section{Results}~\label{sec:experiment}
We apply ASSMT to two datasets: 1) nonstationary simulated data and 2) EEG data from a patient under propofol anesthesia. We compare the spectrogram estimates between MT, SSMT, and ASSMT. All spectrograms are in the dB scale.

\subsection{Application to Simulated Data}
We simulate the data as a superposition of amplitude-modulated process $y_{t,1}$ and frequency-modulated process $y_{t,2}$. The process $y_{t,1}$ is generated from an AR(6) process centered at 11 Hz. The process $y_{t,2}$ is generated from ARMA(6,4) with varying pole loci. The observations are given by $y_t=y_{t,1}\cos(2\pi f_0t)+y_{t,2}+\sigma v_t$, where $f_0=0.02$ Hz, $v_t\sim\mathcal{N}(0,1)$ and $\sigma$ is chosen to achieve an SNR of 30 dB. More details can be found in \cite{Das}.
For MT, we use 6-second windows with 50\% overlap and $M=3$ Slepian tapers, and 6-second non-overlapping windows for both SSMT and ASSMT.
For SSMT, we use the entire data to estimate the parameters. For ASSMT, we use the initial 300 seconds of the data to compute the baseline parameters and use $\alpha=0.95$.

\begin{figure}[!t]
\centering
\includegraphics[width=\linewidth]{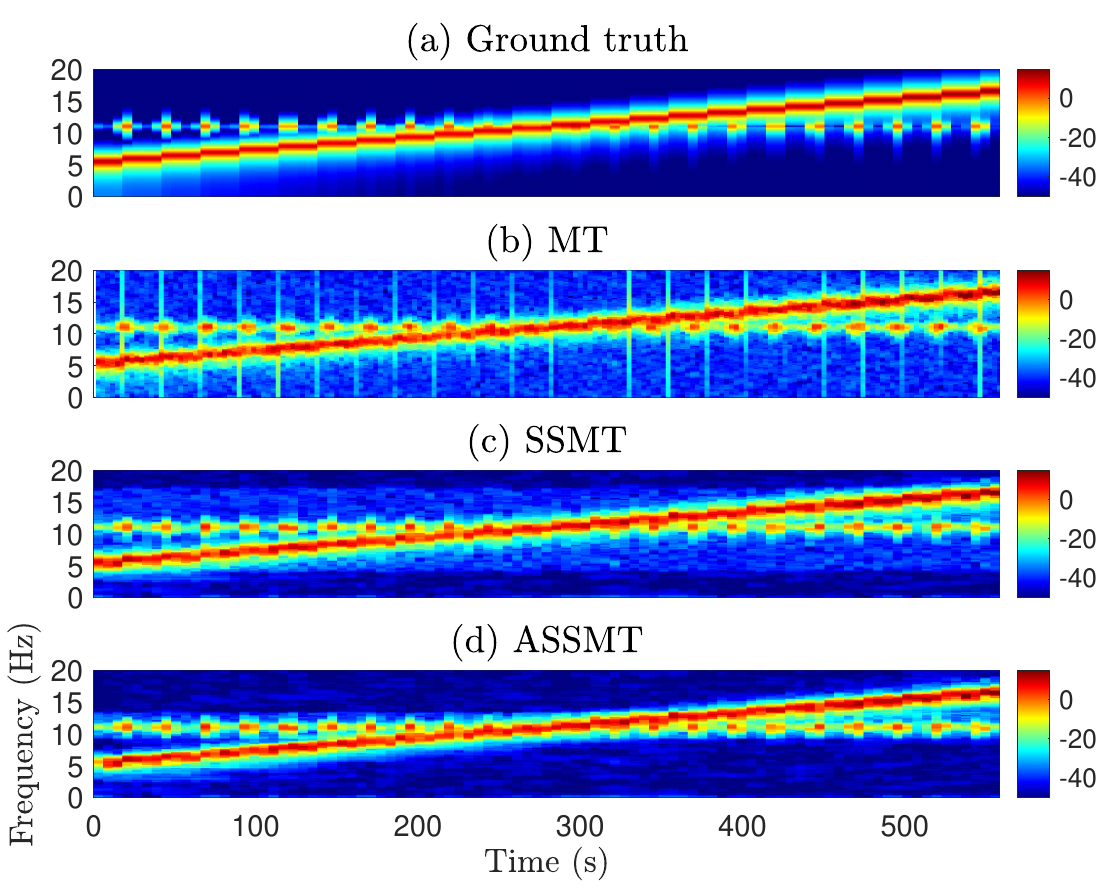}
  \caption{Spectrograms for simulated data (a) ground truth (b) MT (c) SSMT (d) ASSMT with $\alpha=0.95$.}
  \label{fig:sim}
\end{figure}

Fig.~\ref{fig:sim} shows the ground truth and the estimated spectrograms. Although MT captures the spectral dynamics reasonably well, it picks up background noise and spectral artifacts (i.e., vertical lines), and induces mixing of adjacent frequency bands due to low resolution. SSMT (Fig.~\ref{fig:sim} (c)) resolves these issues with sharper spectral localization and removal of spectral artifacts, benefitting from the state-space prior.

ASSMT also shares the artifact rejection and noise reduction properties of SSMT. Moreover, ASSMT performs better denoising, as evidenced in the $5$ to $20$ Hz frequency band. The Itakura-Saito divergence (IS)~\cite{Itakura} between the ground truth and the spectrogram estimate also confirms this observation, with ASSMT attaining the lowest value ($\text{IS}^{\text{MT}}=6.51$, $\text{IS}^{\text{SSMT}}=3.16$, $\text{IS}^{\text{ASSMT}}=\mathbf{2.75}$). We attribute this difference to SSMT's state variance $\widehat{\sigma}_{v,j}^{2,(m),\text{SSMT}}$ and Kalman gain fixed to high values, between $5$ to $20$ Hz.
However, since ASSMT does not commit to a fixed value, it can adaptively change the Kalman gain at different regimes of the data.

\subsection{Application to EEG recorded during anesthesia}
\begin{figure}[!t]
\centering
\includegraphics[width=0.98\linewidth]{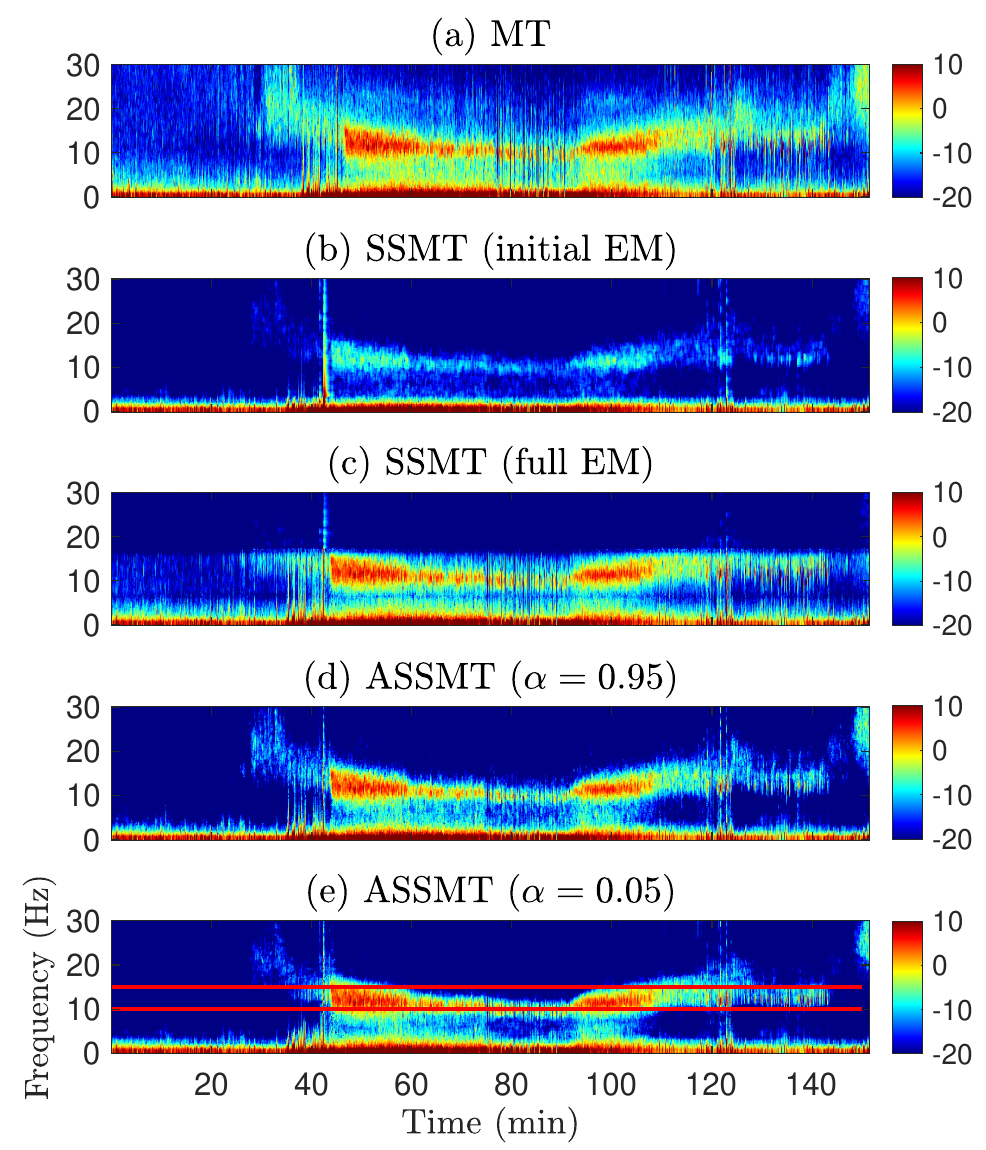}
  \caption{Propofol anesthesia EEG spectrograms (a) MT (b) SSMT with initial 4-min EM window (c) SSMT with full data EM window (d) ASSMT with $\alpha=0.95$ (e) ASSMT with $\alpha=0.05$. Both ASSMT use initial 4-min EM window. Red horizontal lines correspond to 10 and 15 Hz.}
  \label{fig:eeg}
\end{figure}

The EEG was recorded ($f_s=250$ Hz) from a volunteer receiving propofol administered with increasing rate, followed by the decreasing rate~\cite{Purdon}. This setup induces altering states of unconsciousness (or brain states), resulting in varying levels of nonstationarity. We used $M=5$ Slepian tapers, $J=1,000$ samples. For SSMT, we estimate the spectrogram based on the parameters estimated from 1) the initial 4 minutes of data (Fig.~\ref{fig:eeg} (b)) and 2) the entire data (Fig.~\ref{fig:eeg} (c)). For ASSMT, we use the initial 4 minutes to compute the baseline parameters. Fig.~\ref{fig:eeg} shows the estimated spectrograms.

\begin{figure}[!t]
\centering
\includegraphics[width=\linewidth]{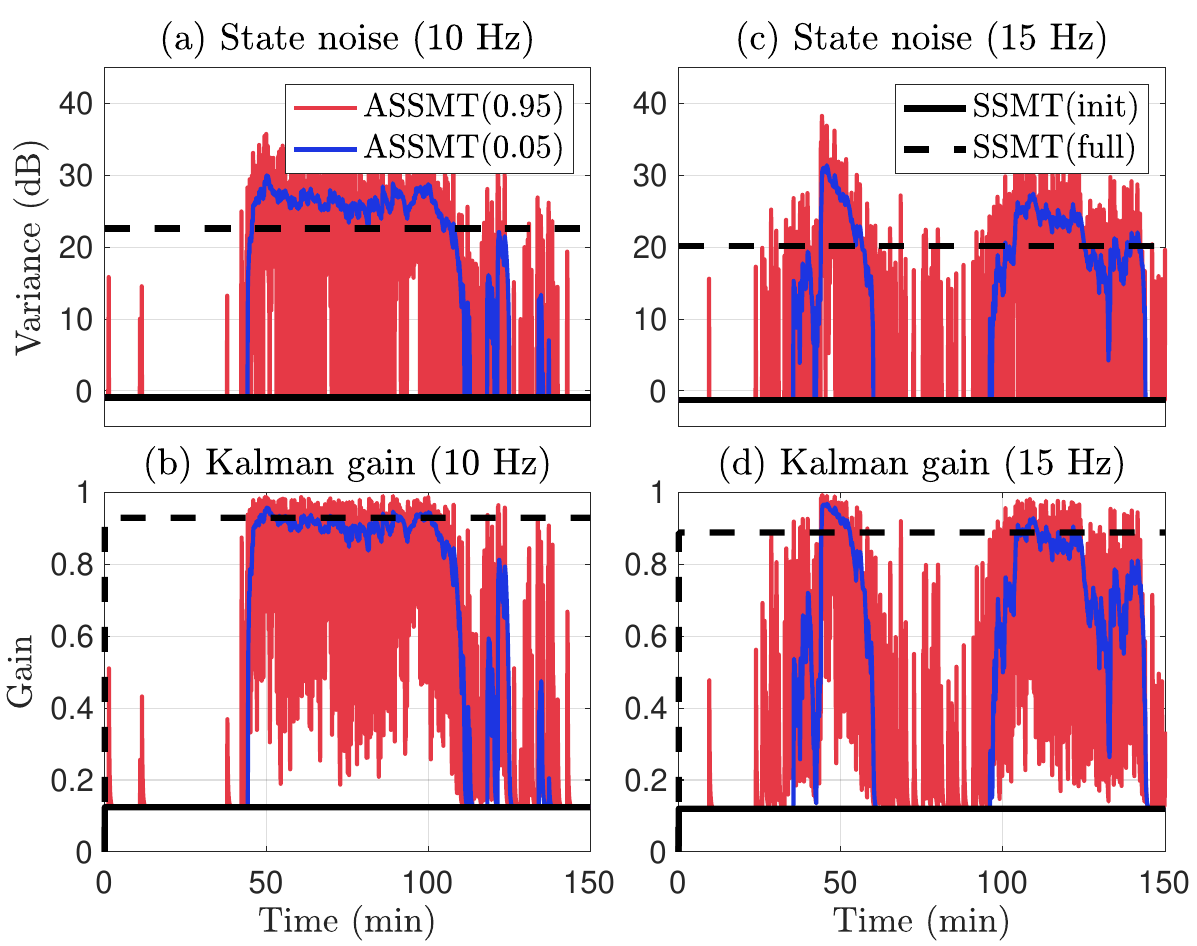}
  \caption{Kalman gain and state noise variance for SSMT and ASSMT for the first taper $m=1$. (a) Kalman gain and (b) State noise variance at 10 Hz. (c) Kalman gain and (d) state noise variance at 15 Hz. }
  \label{fig:KG}
\end{figure}

\noindent \textbf{SSMT vs. ASSMT:}
SSMT based on the initial 4 minutes of the data (Fig.~\ref{fig:eeg} (b)) produces low estimates for $\widehat{\sigma}_{v,j}^{2,(m)}$ due to absence of spectral dynamics for the baseline state. Although it removes background noise well, as a result, it misses most of the strong spectral fluctuation, as evidenced in extreme denoising of the spectra from 40 min to 120 min.

This can be mitigated by applying EM to a different section of the data, or the entire data. Due to high estimates for $\widehat{\sigma}_{v,j}^{2,(m)}$, SSMT better captures the strong nonstationarity (Fig.~\ref{fig:eeg} (c)). However, it fails to denoise the baseline state ($0$ to $40$ min) due to high Kalman gain, as similarly observed in the simulation. These results demonstrate a drawback of the time-invariant assumption, as $\widehat{\sigma}_{v,j}^{2,(m)}$ estimated from different sections could yield significantly different spectrogram estimates.

In contrast, ASSMT adaptively denoises the spectrogram (Fig.~\ref{fig:eeg} (d-e)) even with the initial baseline parameters, the same setting for which SSMT failed (Fig.~\ref{fig:eeg} (b)). The time-varying nature of the model allows it to switch between low $\widehat{\sigma}_{v,k,j}^{2,(m)}$, appropriate for removing background noise, and high $\widehat{\sigma}_{v,k,j}^{2,(m)}$, appropriate for capturing the spectral fluctuation, without fully committing to either parameters. 

Another difference is the computation time. The time for using EM on the entire data and Kalman filtering is $400$ seconds. For ASSMT, however, the procedure takes $6$ seconds, demonstrating an appreciable reduction in computational time and thus, making it more suitable for real-time application. 

\noindent \textbf{Effect of Kalman gain:} To further understand ASSMT, we analyze the evolution of state variance and the Kalman gain across time at the representative frequency bands (10 and 15 Hz), shown in Fig.~\ref{fig:KG}. For both frequency bands, we observe that the state variance and consequently the Kalman gain is increased above the threshold, in tandem with the large spectral fluctuation, as desired. On the other hand, The SSMT Kalman gain stays fixed at either 0.1 (for initial 4-min estimation, solid black in Fig.~\ref{fig:KG}) or 0.9 (for entire data estimation, dotted black in Fig.~\ref{fig:KG}). 

\noindent \textbf{Effect of $\pmb{\alpha}$:} We observe that the denoising performance of ASSMT is robust towards the choice of $\alpha$. To further understand how $\alpha$ affects the filtering/denoising operation, we analyze how $\widehat{\sigma}^{2,(m)}_{v,k,j}$ and Kalman gain change over time. ASSMT with $\alpha=0.95$ (red) is dominated by the heavy fluctuation, $|Y_{k,j}^{(m),F}-Y_{k-1,j}^{(m),F}|^2$. In contrast, ASSMT with small $\alpha=0.05$ (blue) shows smoother state variance and Kalman gain. In this case, ASSMT starts adapting when there is significant evidence for nonstationarity. This explains why $\alpha=0.05$ performs more denoising than $\alpha=0.95$, as it is less susceptible to instantaneous fluctuations (40 min., 0 to 10 Hz) and background noise (75 to 95 min., 2 to 10 Hz).

\section{Conclusion}
We introduced an adaptive state-space multitaper (ASSMT) framework for estimating spectral dynamics in the nonstationary time series. By relaxing the time-invariant parameters assumption and proposing an adaptive parameter estimation scheme, we demonstrated that ASSMT was able to capture strong power fluctuations more reliably compared to SSMT.

\bibliographystyle{IEEEtran}
\bibliography{assmt}

\section*{A. Application to EEG recorded during anesthesia}
We apply ASSMT to EEG recording of another volunteer receiving propofol anesthesia. All parameter settings of SSMT and ASSMT are same as those used in the main paper.

\begin{figure}[!ht]
	\centering
	\includegraphics[width=\linewidth]{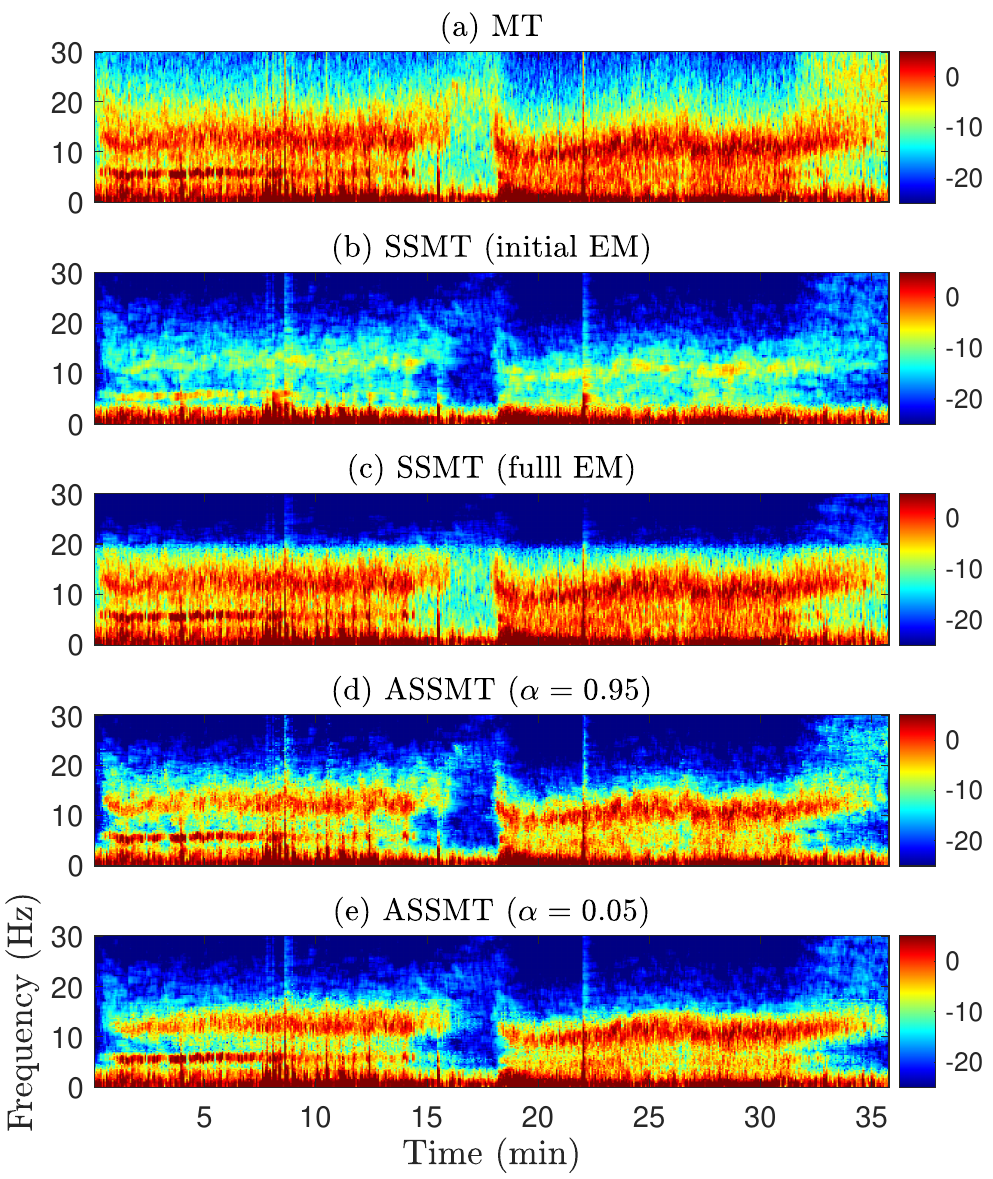}
	\caption{Propofol anesthesia EEG spectrograms (a) MT (b) SSMT with initial 4-min EM window (c) SSMT with full data EM window (d) ASSMT with $\alpha=0.95$ (e) ASSMT with $\alpha=0.05$. Both ASSMT use initial 4-min EM window.}
	\label{fig:eeg2}
\end{figure}

\end{document}